# Tunable mid-infrared hyperbolic van der Waals metasurfaces by strong plasmon-phonon polaritons coupling

*Xueli Wang[1], Kaili Chang[1], Weitao Liu[1], Hongqin Wang[1], Kaihui Liu[2], Ke Chen[1,*]*

[1] Center for the Physics of Low-Dimensional Materials, School of Physics and Electronics, University of Henan, Kaifeng, 475004, China

[2] State Key Laboratory for Mesoscopic Physics, School of Physics, Peking University, Beijing, 100871, China

**ABSTRACT:** Hyperbolic metasurfaces based on van der Waals (vdW) materials support propagation of extremely anisotropic polaritons towards nanoscale light compression and manipulation, and thus has great potential in the applications of planar hyperlens, nanolasing, quantum optics and ultrasensitive infrared spectroscopy. Two-dimensional hexagonal boron nitride (*h*-BN) as a vdW metasurface can manipulate the propagation of hyperbolic polaritons at the level of single atomic layers, possessing higher degree of field confinement and lower losses than the conventional media. However, active manipulation of hyperbolic polaritonic waves in *h*-BN midinfrared metasurfaces remains elusive. Herein, we provide an effective strategy for constructing tunable mid-infrared hyperbolic vdW metasurfaces (HMSs). They are composed of meta-atoms that are the in-plane heterostructures of thin-layer *h*-BN and monolayer graphene strips (iHBNG). The strong coupling of *h*-BN phonons and graphene plasmons enables the large tunability of light fields by tailoring chemical potentials of graphene without frequency shift, which involves topological transitions of polaritonic modes, unidirectional polariton propagation and local-density-of-state enhancement. Simulated visual near-field distributions of iHBNG metasurfaces reveal the unique transformations of hyperbolic polariton propagations, distinguished from that of individual *h*-BN and graphene metasurfaces. Our findings provide a platform of optical nanomanipulation towards emerging on-chip polaritonic devices.

[*]Corresponding author Email: kchen@henu.edu.cn

Hyperbolic metasurfaces (HMSs) are strongly anisotropic flat media whose in-plane effective permittivities along the two axes have opposite signs.[1,2] Extremely anisotropic polaritons along the metasurface possess a hyperbolic in-plan dispersion. This generates a variety of intriguing nanophotonic phenomena, such as deep subwavelength confinement, polarization manipulation, nano-light canalization, and optical super-resolution, *etc*. In metasurfaces, metal plasmonic antennas are generally used as optical scatterers to achieve abrupt phase change and wave-front control by engineering light-mater interactions.[3] Beyond metal media, two-dimensional van der Waals (2D vdW) materials can support more highly confined polariton modes with extremely low losses, thus manipulating light propagation with high quality factors. Graphene plasmonics open the era of 2D polaritonics within atomically thin layers, outperforming traditional plasmonic materials for metasurfaces due to the large tunability.[4] As such, 2D materials beyond graphene, *e.g.*, hexagonal born nitride (*h*-BN) [5] and transition metal dichalcogenides (TMDCs),[6] exhibit extraordinarily large optical refractive indices and binding-energy excitons, thus provide an enticing platform for manipulate light-matter interactions across broadband spectral ranges. Their applications include planar hyperlens,[7] biosensing,[8] optical nano-imaging[9,10] and ultrasensitive infrared spectroscopy.[11,12]

Metasurfaces of 2D crystals are generally divided into two categories. First, naturally existing biaxial vdW crystals, *e.g.*, α-phase molybdenum trioxide (α-MoO$_3$), can achieve the intrinsic in-plane hyperbolic response at the mid-infrared wavelength range.[13,14] Second, artificially patterned uniaxial crystals, generally considered as graphene[15,16] and *h*-BN subwavelength gratings,[17,18] describe a hyperbolic in-plane dispersion with negative permittivity at one axis but positive permittivity at the other axis. For instance, monolayer graphene can be fabricated into nanostructured patterns or periodically arranged strips as plasmonic meta-atoms to create a terahertz HMS.[15,16] Although the hyperbolic in-plane dispersion can be actively modulated, the relatively high loss and short propagating distance of graphene plasmons limit the enhancement of local photonic density of states (LPDOS) in such metasurfaces. Fortunately, phonon polaritons (PhPs) in *h*-BN vdW crystals can extremely confine infrared fields and propagate along surface with notably low losses and picosecond-long lifetime. In contrast to graphene counterparts, thus, *h*-BN slabs also exhibit a topological transition from isotropic to hyperbolic in-plane dispersion upon fabrication into gratings,[17,18] although *h*-BN crystal is a natural out-of-plane hyperbolic material. That is because the effective permittivities are different along the three axes in such a grating structure. Apparently, the *h*-BN hyperbolic metasurfaces bear potential for mid-infrared nanophotonic devices with enhanced light-matter interactions. According to intrinsic nature of phonon and limited geometric tunability, however, actively manipulating hyperbolic curved wavefronts in *h*-BN midinfrared metasurfaces remains elusive.

Many researches show that *h*-BN phonon polaron can be well coupled with graphene plasmon to form hybrid plasmon-phonon polaritons.[19-21] The hybridized polaritons inherit the electrical tunability of graphene plasmon and the long lifetime of *h*-BN phonon.[21] In in-plane graphene/*h*-BN heterointerface, the type of hyperbolic phonon-plasmon polaritons (HP$^3$) almost does not suffer from ohmic losses, enabling a long-distance propagation with the length of ~2 times larger than that of bare *h*-BN layers.[19] It thus provides a promising way to actively manipulate wavefronts in *h*-BN metasurfaces by the strong plasmon-phonon coupling. Herein, we present a strategy for tunable mid-infrared HMSs, which composed of periodic in-plane heterostructures combining *h*-BN thin layers with monolayer graphene strips (iHBNG). High compression and active manipulation of optical field in iHBNG metasurfacs are achieved with large tunability and ultra-low losses by tailoring chemical potentials of graphene. Furthermore, we develop a model of polaritons interactions to illustrate the extreme topological transitions for hyperbolic polariton propagation, demonstrating intuitive and visualized near-field optical images of iHBNG metasurfacs. It offers inspiration for design and applications of emerging on-chip polaritonic devices.

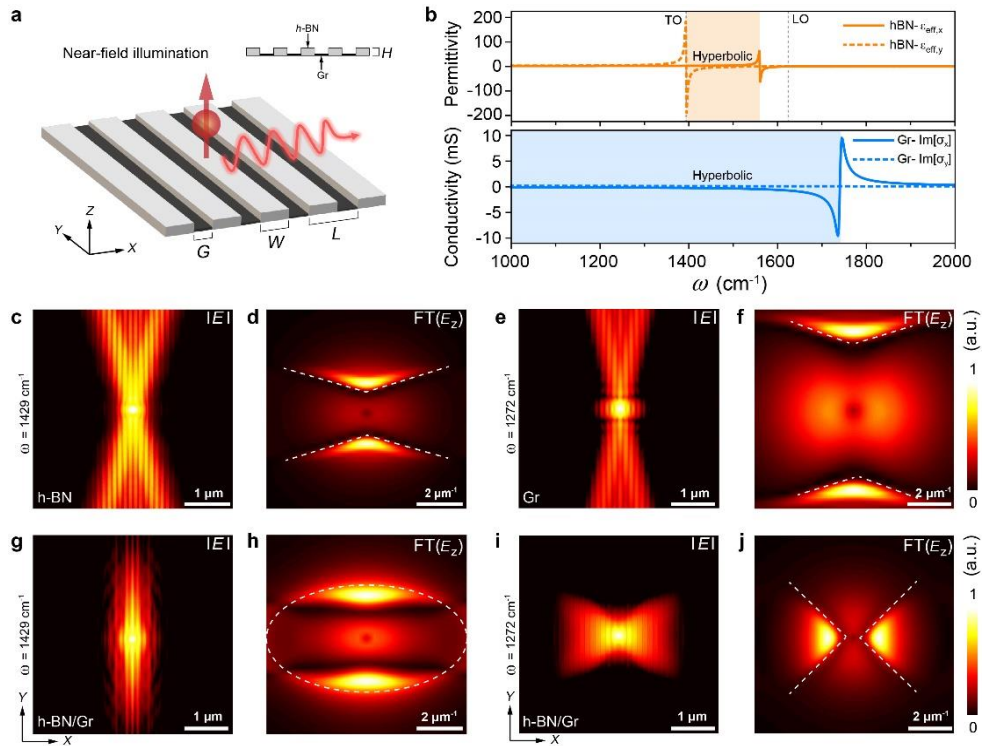

**Figure 1 Topological transition of polaritons in an iHBNG metasurface.** (a) Schematic of a dipole-launched iHBNG metasurface. X-direction, across the *h*-BN

strips; Y-direction, along the *h*-BN strips. (b) Anisotropic effective dielectric permittivities (real parts) of the individual *h*-BN gratings (up, strip width W = 60 nm) and effective conductivity (imaginary parts) of the individual graphene gratings (down, strip width G = 60 nm). (c, e) Simulated magnitude of the near-field distributions above the *h*-BN HMS (1429 cm$^{-1}$, c) and the graphene HMS (1272 cm$^{-1}$, e), respectively, |*E*|. (d, f) Absolute value of the corresponding Fourier transform (FT) of the simulated near-field distributions $E_z$ (**Supplementary Figure 1a** and **1b**). (g, i) Simulated magnitude of the near-field distributions above the iHBNG metasurface with different operation frequencies (g, ω =1429 cm$^{-1}$, graphene chemical potential $\mu_c$ = 0.3 eV; i, 1272 cm$^{-1}$, $\mu_c$ = 0.6 eV). (h, j) Absolute value of the corresponding FT of the simulated near-field distributions $E_z$ (**Supplementary Figure 1c** and **1d**). The period (L= W + G) of iHBNG metasurface is 120 nm in all simulations.

RESULTS AND DISCUSSION

The iHBNG metasurface is composed of thin-layer *h*-BN gratings embedded with monolayer graphene strips in the grating gaps, as schematically shown in **Figure 1a**. In such metasurface, the natural *h*-BN slabs are precisely fabricated into nanogratings as meta-atoms by E-beam lithography on SiO$_2$/Si substrates. While monolayer graphene could be directly grown within the gaps between *h*-BN strips, forming the alternant in-plane heterostructures. To illustrator topological transitions of the iHBNG metasurface, the effective anisotropic permittivities of individual *h*-BN metasurface are first calculated as below according to a modified effective medium model,[17, 18, 22]

$$\varepsilon_{\mathrm{eff},x} = \left(\frac{1-\xi}{\varepsilon_{\mathrm{hBN},\perp}} + \frac{\xi}{\varepsilon_c}\right)^{-1} \qquad (1)$$

$$\varepsilon_{\text{eff},y} = (1-\xi)\varepsilon_{\text{hBN},\perp} + \xi\varepsilon_{\text{air}} \qquad (2)$$

$$\varepsilon_{\text{eff},z} = (1-\xi)\varepsilon_{\text{hBN},\parallel} + \xi\varepsilon_{\text{air}} \qquad (3)$$

$$\xi = \frac{g}{L}, \quad \varepsilon_c = \frac{2L}{\pi h}\ln\left[\csc\left(\frac{\pi}{2}\xi\right)\right] \qquad (4)$$

Where $\xi$ is the filling factor. In this model, the nonlocal effect of strong polaritonic near-field coupling between strips is considered. $\varepsilon_c$ is a nonlocal correction parameter, relying on the period, thickness, and filling factor of nanogratings. Notably, the effect of $SiO_2$/Si substrates on the metasurface is neglected to preserve the intrinsic polariton dispersion relation.[21] In general, the natural h-BN crystal is a polar uniaxial material, exhibiting the out-of-plane hyperbolic dispersion,[5,23] but the in-plane isotropic dispersion.[23] It has two kinds of infrared-active phonon modes with hyperbolicity, and its upper Reststrahlen (RS) band shows type-II hyperbolicity in the frequency range from 1395 to 1630 cm$^{-1}$ ($\varepsilon_\perp < 0$, $\varepsilon_\parallel > 0$). Once the periodic structure is patterned, nevertheless, the permittivities at two in-plane axes are anisotropic, i.e., $\varepsilon_{\perp x} > 0$, $\varepsilon_{\perp y} < 0$, thus the original isotropic polaritons become hyperbolic.

To clarify the photonic phenomena, we calculate the effective permittivity of the h-BN metasurface by the effective medium theory (**Figure 1b**, upper panel).[17,18] The frequency of hyperbolic polaritons ranges from 1395 to 1561 cm$^{-1}$, relying on the filling factor ($\xi = 0.5$) and h-BN strip width (W = 60 nm). As a comparison, we also calculate similarly the in-plane effective conductivity of individual graphene nanogratings, which are complemented with above h-BN nanogratings (**Figure 1b**, lower panel),[15,16]

$$\sigma_{eff} = \begin{pmatrix} \sigma_x & \sigma_{xy} \\ \sigma_{yx} & \sigma_y \end{pmatrix} \qquad (5)$$

$$\sigma_x = \frac{G\sigma\sigma_C}{L\sigma_C + G\sigma} \quad \sigma_y = \sigma\frac{G}{L} \quad \sigma_{xy} = \sigma_{yx} = 0 \quad (6)$$

$$\sigma_C = -i2\omega\varepsilon_0\varepsilon_{eff}(L/\pi)\ln[\csc(\pi W/2L)] \quad (7)$$

Where G is the graphene strip width, W is the gap between adjacent strips (i.e., the width of complementary *h*-BN strips), σ is the graphene conductivity, $\omega$ is the radial frequency, $\varepsilon_0$ and $\varepsilon_{eff}$ are the permittivity of free space and the one relative to the surrounding medium. Note that Equation 7 reveals an effective conductivity $\sigma_C$ related to the nonlocal near-field coupling between adjacent graphene strips. In such metasurface, the *y*-component of effective conductivity is inductive (Im $[\sigma_y^{eff}] > 0$) over the entire mid-infrared frequency band, while the *x*-component of effective conductivity shows a resonance at $\frac{\sigma_C}{W} + \frac{\sigma}{G} = 0$, Im $[\sigma_x^{eff}] < 0$ at frequencies lower than the resonance and Im $[\sigma_x^{eff}] > 0$ at higher frequencies. Because the strong near-field coupling determines the capacitive response of conductivity components at low frequencies. If the signs of Im $[\sigma_x^{eff}]$ and Im $[\sigma_y^{eff}]$ are opposite, an isofrequency surface of hyperbolic dispersion is described. The hyperbolic dispersion of individual graphene metasurface can also change as a function of geometrical and intrinsically electronic parameters of graphene.

To verify geometric designs of above hyperbolic metasurfaces, near-field distribution of dipole-launched polaritons is numerically simulated. The *h*-BN metasurface with strong in-plane anisotropy shows a hyperbolic isofrequency contour (IFC) in *k*-space (**Figure 1c** and **1d**). It reveals that the polaritons are highly confined and propagate anisotropically in a ray-like shape along the surface. The corresponding Fourier transform (FT) describes open hyperboloids of the IFC (**Figure 1d**). Besides, the graphene metasurface also exhibits the similar near-field distribution of hyperbolic polaritons (**Figure 1e** and **1f**). If the above-mentioned two metasurfaces are merged into the iHBNG metasurface, more intriguingly, the polaritons exhibit the distinct dispersion diagrams from the individual hyperbolic metasurfaces (**Figure 1g-1j**). Under the dipole launching with the same frequency, an elliptic dispersion along the Y-axis is presented at $\omega$ = 1429 cm$^{-1}$ (**Figure 1g** and **1h**), while a hyperbolic dispersion along the X-axis is presented $\omega$ = 1272 cm$^{-1}$ (**Figure 1i** and **1j**). Notably, the FT analysis of near-field distributions $E_z$ reveals that the corresponding wave vectors become larger than that of the HP$^2$ on individual *h*-BN metasurfaces (**Figure 1d** and **1h**), while smaller than that of hyperbolic plasmons on individual graphene metasurfaces (**Figure 1f** and **1j**). The above demonstrations reveal either the topological transitions from hyperbolic to elliptical IFC at the hyperbolic frequency of *h*-BN metasurfaces, or the variation of polariton propagation direction from Y- to X-axis at the hyperbolic frequency of graphene metasurfaces.

To study the influence of geometric structure on near-field distribution of individual *h*-BN HMSs, we carry out the numerical simulations (**Supplementary Figure 2**). The dispersion of hyperbolic phonons features two sets of Reststrahlen bands (I and II) in a *h*-BN slab with the thickness of 20 nm, which is visualized as a false-color map of the imaginary part of reflection coefficient r$_p$ (**Figure 2a**).[19] **Figure 2b** shows the variation of nonlocal resonant frequencies revealed by effective permittivity $\varepsilon_{\text{eff,x}}$ in *h*-BN metasurfaces with different strip widths. The resonant frequency shows a red shift with the increase of strip widths from 55 to 70 nm. Meanwhile, the open angle of near-field

IFC, defined by $\theta = \arctan\left(\sqrt{-\frac{\varepsilon_{eff,x}}{\varepsilon_{eff,y}}}\right)$,[24] increases with strip width. Near-field imaging simulations exhibit the extremely anisotropic propagation of in-plane hyperbolic PhPs along the metasurface, *i.e.*, a typical ray pattern of hyperbolic polaritons (**Figure 2c** and **2d**). It reveals that interstrip near-field coupling can give rise to the strong polarization, thus generating the nonlocal effect of hyperbolic dispersion.

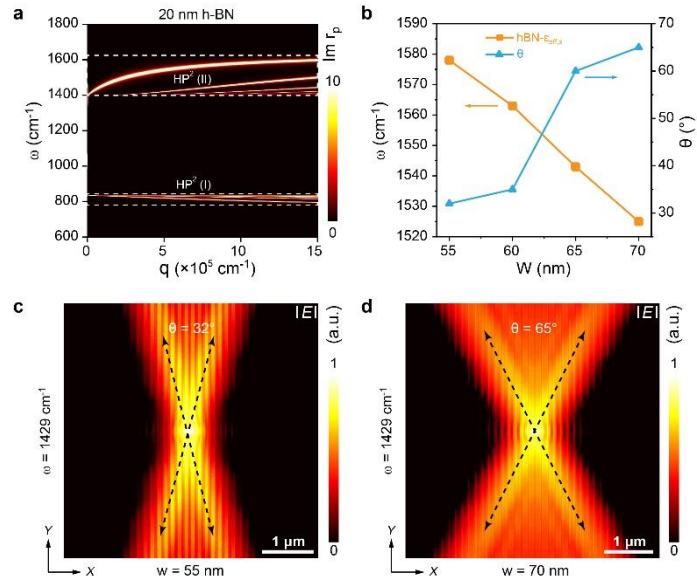

**Figure 2 Geometric-dependent near-field distributions of *h*-BN HMSs.** (a) Calculated dispersion of hyperbolic phonon polaritons (HP$^2$) in *h*-BN slabs with the thickness of 20 nm. (b) Nonlocal resonant frequencies related to effective permittivity $\varepsilon_{eff,x}$ as well as open angles of near-field distribution at different widths of *h*-BN strips. (c, d) Simulated magnitude of near-field distributions in the *h*-BN HMSs, |*E*|. The open angle of the IFC is 32° and 65°, respectively, at W = 55 (c) and 70 nm (d), L = 120 nm.

To realize the strong coupling between *h*-BN phonons and graphene plasmons, a suitable width ratio for *h*-BN and graphene strips is critical for iHBNG metasurfaces (see **Supplementary Figure 3**). **Figure 3a** shows the effect of graphene strip width on the phonon resonant frequency. The frequency related to hyperbolic dispersion becomes larger with the decrease of graphene strip widths. The change of frequency ($\Delta\omega$) is about 375 cm$^{-1}$ in the range of strip widths from 50 to 65 nm. Besides, the open angle of hyperbolic IFC increases with strip widths, analogous to that of the *h*-BN metasurface. As a result, a 1:1 ratio of strip width (60 / 60 nm) for iHBNG metasurfaces is designed to realize minimum coupling losses.

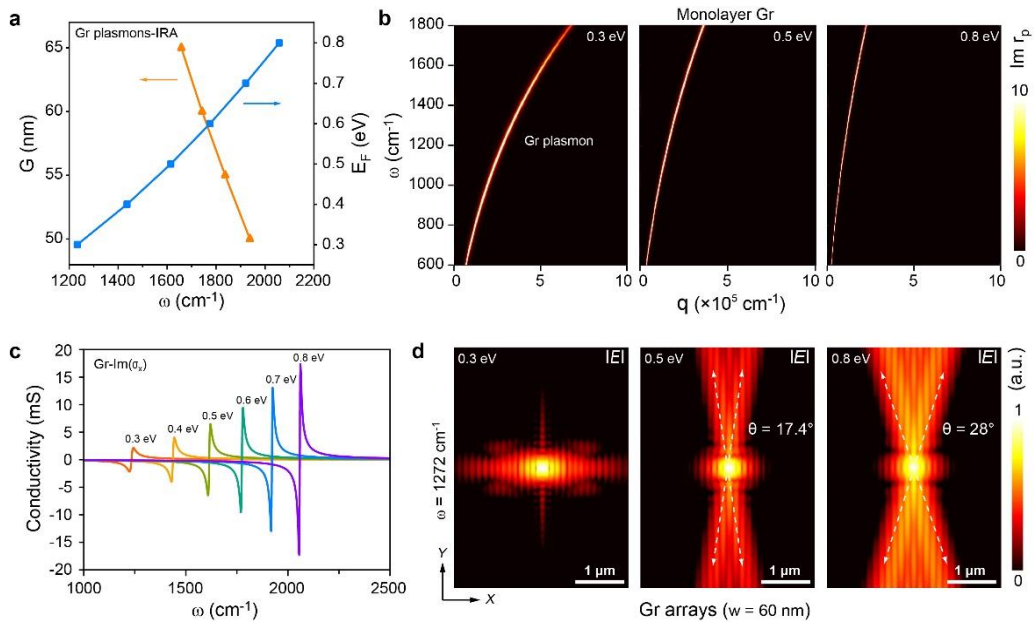

**Figure 3 Chemical potential-dependent tunability of graphene hyperbolic plasmons and their topological transitions.** (a) Resonant frequency change of nonlocal effective optical conductivity Im[$\sigma_x$] under different strip widths and graphene chemical potentials. (b) Calculated dispersion of the surface plasmon polaritons (SP$^2$) of graphene with chemical potentials of 0.3, 0.5 and 0.8 eV. (c) Effective optical conductivities for the graphene metasurface at different chemical potentials. L = 120 nm, G = 60 nm. (d) Simulated magnitude of the near-field distributions above the

graphene metasurfaces with the chemical potentials of 0.3, 0.5 and 0.8 eV, |E|. L = 120 nm, G = 60 nm.

To achieve the tunability of above iHBNG metasurfaces, the plasmons from monolayer graphene strips are modulated by varying chemical potentials of graphene, in order to manipulate the propagation of hybrid polaritons and tailor the local density of states (LDOS) (**Supplementary Figure 4**). The hyperbolic plasmons and topological transitions over graphene gratings are first explored. In general, the dispersion of graphene plasmons can be significantly affected by varying the chemical potential $\mu_e$ (**Figure 3b**). The wave vector $k$ decreases at the same frequency $\omega$, with the increase of $\mu_e$. Furthermore, we carry out theoretical calculation of effective optical conductivity as well as near-field distribution simulation about the graphene metasurface at different $\mu_e$, respectively. With the increase of $\mu_e$, the resonant frequency shows a blue shift, and the range of hyperbolic dispersion frequency becomes broader (**Figure 3c**). Indeed, the variation of such hyperbolic dispersion is related to $\text{Im}\left[\sigma_x^{eff}\right]$ of graphene. Moreover, simulated near-field intensity images show the $\mu_e$-dependent transitions of IFCs from elliptical to hyperbolic (**Figure 3d**), consistent with the analysis of effective optical conductivity. At the same resonant frequency, ~1272 cm$^{-1}$, the real part of effective optical conductivity across the strips is significantly larger than its orthogonal counterpart at $\mu_e$ = 0.3 eV. Where the signs of $\text{Im}\left[\sigma_x^{eff}\right]$ and $\text{Im}\left[\sigma_y^{eff}\right]$ are positive but different absolute values, the propagating plasmonic polaritons exhibit an elliptically-shaped dispersion diagram or low-loss canalization along X direction.[25]

The hyperbolic dispersion diagrams are presented at the elevated $\mu_e$. The open angle of hyperbolic IFCs also increases with $\mu_e$, which is defined by $\theta = \arctan\left(\sqrt{-\frac{\sigma_x}{\sigma_y}}\right)$.[26]

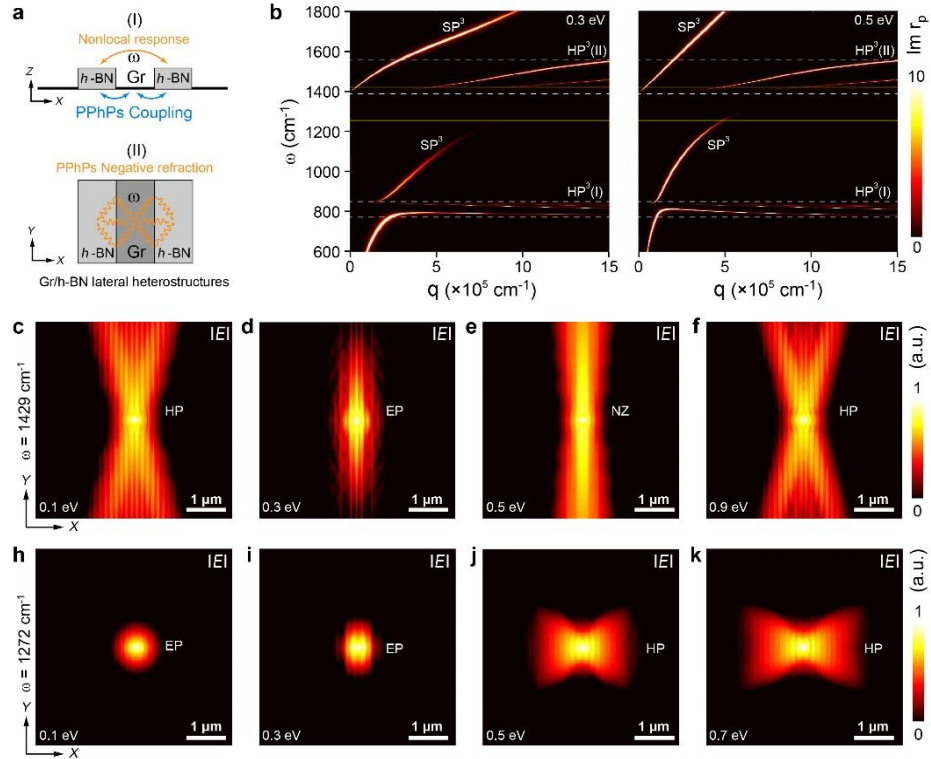

**Figure 4 Tunability of iHBNG metasurfaces**. (a) Schematic of the coupling between graphene plasmons and *h*-BN phonons. (b) Calculated $HP^3$ and $SP^3$ dispersion of the heterostructure of monolayer graphene and thin-layer *h*-BN. (c-f) Simulated magnitude of near-field distributions above the iHBNG metasurfaces at $\omega = 1429$ cm$^{-1}$ with different chemical potentials, $|E|$. $\mu_e$ = 0.1, 0.3, 0.5, and 0.9 eV. (g-j) Simulated magnitude of near-field distributions above the iHBNG metasurfaces at $\omega = 1272$ cm$^{-1}$ with different chemical potentials, $|E|$. $\mu_e$ = 0.1, 0.3, 0.5, and 0.7 eV.

Considering the tunability of aforementioned graphene plasmons, we can manipulate the light field of iHBNG metasurfaces by the resonance matching between *h*-BN phonons and graphene plasmons. At the frequency range of *h*-BN hyperbolic dispersion, the embedded graphene plasmons in *h*-BN metasurface can extremely enhance the nonlocal response between adjacent *h*-BN strips (**Figure 4a**). This can be revealed by the calculated dispersion of *h*-BN hyperbolic phonons coupled with graphene plasmons (**Figure 4b**, **Supplementary Figure 5** and **Note 1**). Hybrid polariton coupling dramatically enhances at the elevated $\mu_e$. Simulated near-field distribution of iHBNG metasurfaces at 1429 cm$^{-1}$ reveals the effect of $\mu_e$ on extreme light confinement and propagation. Intriguingly, a new mode of light field is presented if the $\mu_e$ is elevated from 0.1 to 0.3 eV (**Figure 4c** and **4d**). When the $\mu_e$ is 0.1 eV, the near field distribution of iHBNG metasurface exhibits an in-plane hyperbolic dispersion at 1429 cm$^{-1}$, analogous to that of the individual *h*-BN metasurface. This is because the plasmons of graphene strips at relatively low $\mu_e$ is difficult to be launched due to the low carrier density. Only few free carriers generate with intra-band transition in such frequency range of incident light.[27] Meanwhile, the momentum difference between graphene plasmons ($k \propto \omega_{pl}^2/E_F$ [28]) and *h*-BN HP$^2$ is very large at low $\mu_e$ under the same frequency, [29] thus the interaction is very weak between the two polaritons. With the increase of $\mu_e$, the momentum of graphene plasmons approaches gradually the *h*-BN HP$^2$ momentum, thus leading to the strong coupling to generate hybrid plasmon-phonon polaritons (**Figure 4b**). In this context, the effective permittivity along X-axis could decrease even have opposite signs (*i.e.*, $\varepsilon_{eff-x} < 0$) in the iHBNG metasurface at 1429 cm$^{-1}$, compared with that of individual *h*-BN metasurfaces ($\varepsilon_{eff-x\ hBN} > 0$). As a result, a topological transition from hyperbolic to elliptical for polariton propagation would be presented without launching frequency shift (**Figure 4d**). The continuous increasement of $\mu_e$ could also give rise to a near-zero mode (NZ) due to the distinct difference between $\varepsilon_{\text{eff-x}}$ and $\varepsilon_{\text{eff-y}}$ (**Figure 4e**). Either so-called canalization modes or near-zero state could be ascribed from the strong near-field coupling and interstrip nonlocal response.[25] If the $\mu_e$ increases up to 1 eV, however, the momentum difference between graphene plasmons and h-BN HP$^2$ enlarge again at the same frequency, thereby the large momentum mismatch results in the weak interactions between such two polaritons. As a result, the near-field distribution of iHBNG metasurfaces describe a similar hyperbolic IFC to individual h-BN and graphene hyperbolic metasurfaces (**Figure 4f**).

Beyond the frequency range of h-BN RS band, more intriguingly, an extremely anisotropic topological transition for iHBNG metasurfaces is presented with the rising of $\mu_e$ at 1272 cm$^{-1}$ (**Figure 4g-j**). The propagation direction of such hyperbolic polaritons is along X-axis (**Figure 4h-j**), although the hyperbolic polaritons propagate along Y-axis in individual graphene metasurfaces at the same frequency (see **Figure 3d**). In this context, the graphene plasmons can couple with the h-BN surface phonons to form surface plasmon-phonon polarons (SP$^3$),[19] thus demonstrating a hyperbolic propagation wavefront. This phenomenon is in accord with the dispersion calculation of h-BN surface phonons coupled with graphene plasmons at high momentum (**Figure 4b**). Moreover, the change of propagation direction could be related with the negative refraction of in-plane polaritons (see **Figure 4a**).[30] The lateral interface between h-BN and graphene strips could support the in-plane negative refraction for high-$k$ polaritons at subwavelength scales.

CONCLUSIONS

In conclusion, we have proposed a hyperbolic midinfrared vdWs metasurface based on the in-plane heterostructures of thin-layer *h*-BN and monolayer graphene strips. The large tunability of nanolight fields in iHBNG metasurfaces is achieved by tailoring hybrid phonon-plasmons polaritons. A topological transition for polariton propagation from hyperbolic to elliptical is demonstrated by the stimulated near-field images, based on tailoring chemical potentials of graphene without launching frequency shift. Besides, the LPDOS enhancement and the propagation direction change of hyperbolic polaritons in iHBNG metasurfaces are also achieved by this approach. These phenomena could be ascribed from the tunable and strong polaritons coupling. This work provides a platform for developing on-chip nanophotonic devices including nanolight modulators, low-loss waveguides, and ultrasensitive molecular detection.

METHODS

**Material Model.** The *h*-BN slab is a polar vdW material with uniaxial dielectric constant and has natural hyperbolic properties[5,23]. It is a vdW crystal with two kinds of infrared-active phonon modes relevant to hyperbolicity, where the lower RS band corresponds to type-I hyperbolicity ($\varepsilon_{//} < 0$, $\varepsilon_{\perp} > 0$) and the upper RS band shows type-II hyperbolicity ($\varepsilon_{\perp} < 0$, $\varepsilon_{//} > 0$). As a result, the phonon polaritons in natural *h*-BN exhibit an out-of-plane hyperbolic dispersion, whereas the in-plane dispersion is isotropic. [23] The permittivity of *h*-BN slabs can be described using the following equation[29],

$$\varepsilon_{hBN,j} = \varepsilon_{\infty,j}\left(1 + \frac{\omega_{LO,j}^2 - \omega_{TO,j}^2}{\omega_{TO,j}^2 - \omega^2 - i\omega\Gamma_j}\right) \quad (8)$$

Where j=⊥ and ∥, out-of-plane $A_{2u}$ phonon modes are $\omega_{TO,\parallel} = 785 \text{ cm}^{-1}$ and $\omega_{LO,\parallel} = 845 \text{ cm}^{-1}$, and the in-plane $E_{1u}$ phonon modes are $\omega_{TO,\perp} = 1395 \text{ cm}^{-1}$ and $\omega_{LO,\parallel} = 1630 \text{ cm}^{-1}$, The other parameters are $\varepsilon_{\infty,\parallel} = 2.8$, $\Gamma_\parallel = 1 \text{cm}^{-1}$, $\varepsilon_{\infty,\perp} = 3$ and $\Gamma_\perp = 2\text{cm}^{-1}$.

Optical response of graphene can be described by optical conductivity according to local random phase approximation. By using Kobo formula,[28] which satisfies the requirement of $K_BT \ll E_f$ at room temperature (T = 300 K), the typical optical conductivity of monolayer graphene is given by:

$$\sigma = \frac{ie^2 E_f}{\pi\hbar^2(\omega + i\tau^{-1})} + \frac{ie^2}{4\pi\hbar}\ln\left[\frac{2|E_f| - \hbar(\omega + i\tau^{-1})}{2|E_f| + \hbar(\omega + i\tau^{-1})}\right] \quad (9)$$

The first term in Eq. 9 represents intra-band transition contribution, and the second term is the contribution of inter-band transition to the total conductivity. Considering the lower energy of incident light in mid-infrared frequency, the contribution of intra-band transition to $\sigma$ dominates. In this formula, $\tau$ is the electron relaxation time, $\hbar$ is the reduced Plank constant, $\omega$ is the angular frequency, and $e$ is electron charge. The mobility and relaxation time are related by $\tau = \mu E_f / ev_f^2$, where μ = 10,000 cm² V⁻¹ s⁻¹ is the carrier mobility and $V_f = 10^6$ m s⁻¹ is the Fermi velocity for graphene. The specific dielectric function of graphene is calculated in **Supplementary Figure 6.**

**Numerical Simulations.** The finite-difference time-domain (FDTD) method was used in all full-wave simulations, based on commercially available software (Ansys-Lumerical-FDTD, 2020 R2). For all indicated frequencies, we simulated the near-field distributions above the sample surfaces. An electric dipole source was placed at 200 nm above the individual $h$-BN, graphene and iHBNG metasurfaces, where $h$-BN strip width W = 60 nm, layer thickness h= 20 nm, and graphene strip width G = 60 nm. The Fourier transformed images were obtained from the simulated near-field distributions at a height of 20 nm above all the metasurfaces.

ASSOCIATED CONTENT

**Supporting Information**

The supporting Information is available free of charge at https://

AUTHOR INFOMATION

**Notes**

The authors declare no competing financial interest.

ACKNOWLEDGMENTS

This project is supported by the National Natural Science Foundation of China (U1904193), the Special Program for Basic Research in University of Henan Province, China (20zx010), the Training Plan of Young Backbone Teachers in Colleges and Universities of Henan Province, China (2019GGJS025), the Science and Technology Development Project of Henan Province, China (212102210454), the Program for Innovative Research Team in Science and Technology in the University of Henan Province (20IRTSTHN012), the Zhongyuan Thousand Talents Program of Henan Province, and the National Young Top-NotchTalents of Ten-Thousand Talents Program.